**Two superconducting phases and their characteristics in layered BaTi$_2$(Sb$_{1-x}$Bi$_x$)$_2$O with x=0.16**[*]

WU Yue(伍岳), DONG Xiao-Li(董晓莉) [**], MA Ming-Wei(马明伟), YANG Huai-Xin(杨槐馨), ZHANG Chao(张超), ZHOU Fang(周放), ZHOU Xing-Jiang(周兴江), ZHAO Zhong-Xian(赵忠贤)[**]

*Beijing National Laboratory for Condensed Matter Physics, Institute of Physics, Chinese Academy of Science, Beijing 100190, China*

Two correlated superconducting phases are identified in layered superconductor BaTi$_2$(Sb$_{1-x}$Bi$_x$)$_2$O (x=0.16), with the superconducting transition temperatures of $T_C$ = 6 K (the high $T_C$ phase) and 3.4 K (the low $T_C$ Phase), respectively. The 6 K superconducting phase appears firstly in the as-prepared sample and can decay into the low $T_C$ phase by exposing to ambient atmosphere for certain duration. Especially the high $T_C$ phase can reappear from the decayed sample with the low $T_C$ phase by vacuum annealing. It is also found that the CDW/SDW order occurs only with the 6 K superconducting phase. These notable features and alteration of superconductivity due to the post-processing and external pressure can be explained by the scenario of electronic phase-separation.

PACS: 74.81.-g, 74.70.Xa, 74.25.-q, 71.45.Lr

[*] supported by projects 11274358 & 11190020 of NSFC and the National Basic Research Program of China (2013CB921700 & 2011CB921703).
[**] Corresponding author. Email: dong@aphy.iphy.ac.cn; zhxzhao@aphy.iphy.ac.cn

Titanium oxypnictide $ATi_2P_2O$ (A=Ba, Na and P=Sb, As) has been proposed as a notable candidate of new superconductor since it shows obvious charge- or spin-density wave (CDW/SDW) anomaly and shares the similar layered structure with the iron-arsenide [1-6]. The recently realized superconductivity up to 6K by the A-site doping or P site substitution in this system continues to arouse further interests [7-17]. Electronic structure calculations of $ATi_2P_2O$ by FLAPW–GGA [13] suggest that three Ti 3d electronic bands near Fermi surface could be involved in the formation of superconducting state. However, it is not clear whether and how CDW/SDW ordering correlates with superconductivity in this family [9]. Besides, the argument that the 6 K superconductivity in $BaTi_2(Sb_{1-x}Bi_x)_2O$ may come from possible Bi impurity due to quenching [17] also needs to be clarified. In this paper, we will report on observations of two superconducting (SC) phases, i.e. with $T_C$ = 6 K (the high $T_C$ phase) and 3.4 K (the low $T_C$ Phase) respectively, identified in $BaTi_2(Sb_{0.84}Bi_{0.16})_2O$. Our experimental data also demonstrate that the 6 K superconducting phase appears firstly in the as-prepared sample and can decay into the low $T_C$ phase by exposing to ambient atmosphere for certain duration. Especially the high $T_C$ phase can reappear from the decayed sample with low $T_C$ phase by vacuum annealing. We also find that the CDW or SDW ordering below 44 K is related to the 6 K superconducting phase, not to the 3.4 K one. The magnetic measurements under high pressure reveal certain notable feature of the Meissner signal of the 3.4 K SC phase. These notable features and alteration of superconducting state observed in the present system can be fundamentally explained by the picture of electronic phase-separation in analogy with what observed in cuprate superconductors.

Polycrystalline samples with the nominal composition of $BaTi_2(Sb_{1-x}Bi_x)_2O$ were synthesized using conventional solid state reaction method. The starting materials BaO (Alfa Aesar, 99.5%), Ti (Alfa Aesar, 99.9%), Sb (Alfa Aesar, 99.9%) and Bi (Alfa Aesar, 99.5%) with chemical stoichiometry were ground and pressed into pellets in the argon-filled glove box. Then the mixtures were sealed in evacuated quartz tubes and heated at 950°C for 36 hours, followed by cooling to room temperature at 50K/hour. Samples' structures were characterized by powder x-ray diffraction (XRD) at room temperature using an 18KW MXP18A-HF diffractometer. Scanning electronic microscopy (SEM) and energy dispersive x-ray spectroscopy (EDXS) analysis were performed by using a Philips XL30 S-FEG microscope. The temperature dependence of resistivity was measured by a Quantum Design physical property measurement system (PPMS-9). Magnetic properties under ambient and high pressure were determined on a Quantum Design magnetic property measurement system (MPMS XL-1). The hydrostatic pressure was generated using a commercial pressure cell (Mcell 10), with the pressure in situ monitored by the $T_C$ shift of a lead manometer [18]. The superconducting transition of each sample was measured down to 1.8 K under an applied field of 1 Oe, while the CDW/SDW transition was measured under 1 Tesla.

Fig. 1(c) shows the powder X-ray diffraction patterns for our samples with various exposure durations. Fundamentally, all main reflection peaks can be well assigned to the reported tetragonal structure with a space group of P4/mmm and lattice parameters of a=4.113(1) Å and c=8.093(1) Å. In addition, we can also see a weak reflection peak from Bi impurity in the diffraction patterns. Given in fig. 2 (a) is the temperature dependent resistivity of the as-prepared $BaTi_2(Sb_{1-x}Bi_x)_2O$ (x=0.16) sample. A clear superconducting transition at 6 K and a CDW/SDW anomaly around 44K confirm the coexistence of superconductivity and CDW/SDW ordering [17]. Both zero-field cooling (ZFC) and field-cooling (FC) magnetic susceptibilities (fig. 2(b)) again verify a single-phased superconducting transition at 6 K in the as-prepared sample.

First of all, it is necessary to clarify that the superconductivity at 6 K in present system is an intrinsic property and does not come from Bi impurity in $BaTi_2(Sb_{1-x}Bi_x)_2O$ sample. Fig. 1 (a) and (b) show SEM images illustrating the microstructure and local chemical composition in our superconducting samples. Fig. 1(a) represents the typical texture features of our samples to the most extent, where no Bi impurity can be detected. On the other hand, it is found at certain grain boundaries that the Bi impurity appears in very limited local area as indicated in fig. 1(b), but neither amorphous Bi coated grains nor amorphous Bi network have been observed. Importantly, the 6 K SC phase can be easily recovered from a well decayed sample with the low $T_C$ SC phase by annealing in vacuum at 800°C, as will be shown in following context. All these facts can definitely rule out the possibility that the superconductivity at 6 K might come from amorphous Bi [17] since the condition for bismuth to become superconducting is quite stringent [19, 20]. Besides , a possible superconducting $BaBi_3$ impurity with Tc of 5.69K[21] is also ruled out from our observation. One reason is that a weak impurity peak at 30 degree（the trace of BaBi3）in the powder X-ray diffraction pattern is only detectable in a few samples, not all of samples with superconductivity at 6 K. The other reason is that $BaBi_3$ impurity is sensitive and easy to decomposes in air [22]. As shown in fig 1c, a tiny impurity peak appears around 30 degree for the sample exposure about 0.5 hour (represented by square). However, this peak disappears when exposed in air for 1 hour (circle) and above. While the 6K superconducting phase is still dominant (figure 3 a, circle). Therefore, the two superconducting phases and their behaviours presented in this work are the intrinsic characteristics of $BaTi_2(Sb_{0.84}Bi_{0.16})_2O$. Moreover, we have also carried out a series of measurements to investigate the alternation of CDW/SDW ordering accompanying with the decay of superconductivity, when the sample is exposed to ambient atmosphere. Fig. 3(a) shows the temperature dependent FC magnetic susceptibilities of an as-prepared $BaTi_2(Sb_{0.84}Bi_{0.16})_2O$ sample undergoing various exposure durations. The change in superconductivity is clearly visible: the single-phased SC transition at 6 K, so-called the high $T_C$ phase, decays to a single-phased SC transition at 3.4 K, so-called the low $T_C$ phase, after one day exposure. What is more interesting is that, only after an exposure of 0.5 hour, the low $T_C$ phase emerges and coexists with the high $T_C$ phase, as can be seen from both magnetic susceptibility and resistivity (lower inset in fig. 3(a)) curves. Moreover, as shown in the upper inset of fig. 3(a), the CDW/SDW ordering recognized by the change in magnetization is also fading gradually during the exposure and finally vanishes accompanying with the disappearance of the SC transition at 6 K, even though the low $T_C$ phase still survives. These facts indicate that the CDW/SDW ordering is essentially related to the high $T_C$ phase rather than the low $T_C$ one. It is worth to note that the decay in superconductivity as well as in CDW/SDW ordering with the exposure is of electronic origin rather than structural one. As can be seen from the XRD data shown in fig. 1(c), no extra peaks are detected and all the peak positions remain unchanged, indicating that no structural corruption or detectable lattice changes occur during the whole exposure process.

As a measure of superconducting volume fraction, the Meissner signal size (FC susceptibility data at given temperature intervals) can be used to semi-quantitatively illustrate the alteration of the superconducting states. Fig. 4 displays the Meissner signal sizes derived from the susceptibility data given in fig. 3(a) and (c) for the superconducting transitions under ambient and high pressure. The definitions of the Meissner signal sizes used in present study are illustrated in the caption of fig. 4.

Several interesting features can be seen from fig. 4. In general the Meissner signal size (data symbolized by hollow blue squares) decreases monotonically with the exposure (a broken x-axis is used in fig. 4), from 13.9% after 0.5 hour exposure to less than 2.3% after one day exposure. Although the onset $T_C$ of the high $T_C$ phase remains almost unchanged until it is exposed for 4 hours (see Fig. 3(a)), its Meissner signal size (hollow brown circles) is found to decrease more rapidly. In contrast, however, the low $T_C$ phase exhibits quite a different behavior. First, as mentioned above, the SC transition at 3.4 K emerges immediately at the initial exposure and coexists with the original SC transition stabilized at 6 K. Second, its Meissner signal size (hollow brown triangles) increases steadily following with the exposure time, accompanied by the decrement in Meissner signal size of the high $T_C$ phase (see the light gray area in fig. 4). It is only after the high $T_C$ phase being smeared out that the Meissner signal size of the low $T_C$ phase starts to decrease, but with its onset $T_C$ remaining at 3.4 K for the case of one day exposure (see both fig. 3(a) and fig. 4). As an important fact, the high $T_C$ phase can reappear from this well decayed sample (with the low $T_C$ phase) after annealing in vacuum at 800℃ for 4 hours (fig. 3(b)). The superconductivity in well decayed sample is also checked by measurements under pressure, where no change in chemical composition is involved. While the $T_C$ of well decayed sample is raised to some extent with pressure, its Meissner signal size (hollow red stars) increases remarkably, over 10 times from 0.64% under 0 GPa to about 6.6% under 0.83 GPa, as illustrated in fig. 3 (c) and fig.4. This phenomenon is indeed remarkable, considering that the applied pressure is mild (only up to 0.83 GPa) and the sample grains inside the pressure cell are spread in the pressure-transmitting medium. It is never observed in single-phased superconducting systems of either conventional type II superconductor like $MgB_2$[23] or iron-based high-temperature superconductors [18]. Instead, it is quite similar to the behaviour of phase-separated cuprate superconductors, where the diamagnetic susceptibilities of two coexistent superconducting phases tuned by pressure were also found to be correlated [24]. With the electronic phase-separation scenario that areas with different charge densities coexist, our observations can be easily understood. When the carrier density in non-superconducting carrier-poor area is increased by external pressure, the carrier-rich area that supports superconductivity can expand dramatically by permeating and spreading into carrier-poor area, as in the case of cuprate superconductors [24-26]. Based on our experimental results, marked in light gray in fig. 4 is the superconducting phase-separation region where coexist two correlated meta-stable SC phases. Outside the phase-separation regime shown in fig.4, the single-phased superconductivity at 6 K only appears in as-prepared sample in the left, and in the right part the remnant single-phased superconductivity at 3.4 K starts to decay in both its SC volume fraction and $T_C$.

As a comparison, shown in the inset of fig. 3(c) are the magnetic susceptibility data under pressures for the high $T_C$ phase. A decrease of $T_C$ with pressure is observed. This is consistent with the previous reported data on pressure effect [9]. The coexistence of CDW/SDW ordering and superconductivity in $BaTi_2(Sb_{1-x}Bi_x)_2O$ has also been studied by other groups. However some issues are still in debate. Zhai et al. [17] reported that CDW/SDW ordering and superconductivity coexist with x up to optimal doping level, while Yajima et al. [16] stated that there was no CDW/SDW anomaly when $x \geq 0.1$. These disagreements in their data can be addressed based on our observations and the phase-separation picture. Even though the doping levels are similar or close, the degree of decay or phase-separation nature of their samples may be quite different. The sample with x being about 0.1 of the latter report seems to be well decayed as revealed by its very small Meissner signal [16], thus no CDW/SDW is detectable.

Our observations shown above indicate that the density and redistribution of charge carrier in BaTi$_2$(Sb$_{1-x}$Bi$_x$)$_2$O can be easily tuned by sample post-process, such as exposing to atmosphere and annealing in vacuum, and by external pressure. We therefore propose that the mixed valence bands of Ti$^{3+}$ and Ti$^{4+}$ near the Fermi surface may play the key role. It deserves further investigations to get insight into the interplay of the mixed valence states of Ti$^{3+}$ and Ti$^{4+}$, CDW/SDW order and superconducting phase-separation. It is also interesting to address the interplay of the high $T_C$ SC phase and CDW/SDW ordering.

We identify two meta-stable and correlated superconducting phases, with $T_C$'s being of 6 K and 3.4 K, respectively, in layered superconductor BaTi$_2$(Sb$_{1-x}$Bi$_x$)$_2$O (x=0.16) exposed to ambient atmosphere for certain duration. The correlation between the two SC phases is manifested as follows: (a) the low $T_C$ phase is derived out of the single high $T_C$ phase in as-prepared sample with the exposure, at the expense of SC volume fraction of the latter; and (b) the high $T_C$ phase can reappear from well decayed sample with the low $T_C$ phase by annealing in vacuum. We find a strong pressure effect on Meissner signal size of the low $T_C$ phase. These observations can be well explained by analogy with the electronic phase-separation in cuprates. It is also found that the CDW/SDW ordering is related to the high $T_C$ SC phase rather than the low $T_C$ one. Our experimental results indicate that the density and redistribution of charge carrier in BaTi$_2$(Sb$_{1-x}$Bi$_x$)$_2$O are very sensitive to sample post-processing, as exposed to atmosphere and annealed in vacuum, and to external pressure. It is proposed that the mixed valence bands of $Ti^{3+}$ and $Ti^{4+}$ near the Fermi surface should be responsible to the observations.

The authors would like to thank Prof. J.Q. Li for valuable discussions.

**References**

[1] Adam A and Schuster H U 1990 Anorg. Allg. Chem. **584** 150

[2] Axtell E A, Ozawa T, Kauzlarich S M and Singh R R P 1997 J. Solid State Chem. **134** 423

[3] Wang X F, Yan Y J, Ying J J, Li J Q, Zhang M, Xu N and Chen X H 2010 J. Phys. Cond. Matter **22** 075702

[4] Singh D J 2012 New J. Phys. **14** 123003

[5] Shi Y G, Wang H P, Zhang X, Wang W D, Huang Y and Wang 2013 N L Phys. Rev. B **88** 144513

[6] Yan X W and Lu Z Y 2013 J. Phys. Cond. Matter **25** 365501

[7] Doan P, Gooch M, Tang Z J, Lorenz B, Moller A, Tapp J, Chu C W and Guloy A M 2012 J. Am. Chem. Soc. **134** 16520

[8] Yajima T, Nakano K, Takeiri F, Ono T, Hosokoshi Y, Matsushita Y, Hester J, Kobayashi Y and Kageyama H 2012 J. Phys. Soc. Jpn. **81** 103706

[9] Gooch M, Doan P, Lorenz B, Tang Z J, Guloy A M and Chu C W 2013 Supercond. Sci. Tech. **26** 125011

[10] Kitagawa S, Ishida K, Nakano K, Yajima T and Kageyama H 2013 Phys. Rev. B **87** 060510

[11] Nakano K, Yajima T, Takeiri F, Green M A, Hester J, Kobayashi Y and Kageyama H 2013 J. Phys. Soc. Jpn. **82** 074707

[12] Pachmayr U and Johrendt D 2013 arXiv:1308.5660v1

[13] Suetin D V and Ivanovskii A L 2013 J. Alloys Compd. **564** 117

[14] Rohr F V, Nesper R and Schilling A 2013 arXiv:1311.1493

[15] Yajima T, Nakano K, Takeiri F, Hester J, Yamamoto T, Kobayashi Y, Tsuji N, Kim J, Fujiwara A and Kageyama H 2013 J. Phys. Soc. Jpn. **82** 013703

[16] Yajima T, Nakano K, Takeiri F, Nozaki Y, Kobayashi Y, Kageyama H 2013 J. Phys. Soc. Jpn. **82** 033705

[17] Zhai H F, Jiao W H, Sun Y L, Bao J K, Jiang H, Yang X J, Tang Z T, Tao Q,

Xu X F, Li Y K, Cao C, Dai J H, Xu Z A and Cao G H 2013 Phys. Rev. B **87** 100502

[18] Dong X L, Lu W, Yang J, Yi W, Li Z C, Zhang C, Ren Z A, Che G C, Sun L L, Zhou F, Zhou X J and Zhao Z X 2010 Phys. Rev. B **82** 212506

[19] Brandt N B and Ginzburg N I 1965 Soviet Physics Uspekhi-Ussr **8** 202

[20] Vi Petrosya, Molin V N, Vasin O I, Pa Skripkin, Stenin S I and Batyev E G 1974 Zhurnal Eksperimentalnoi I Teoreticheskoi Fiziki **66** 996

[21] Matthias B T and Hulm J K 1952 Phys. Rev. **87** 799

[22] Saparov B and Sefat A S 2013 J. Solid State Chem. **204** 32

[23] Tomita T, Hamlin J, Schilling J, Hinks D and Jorgensen J 2001 Phys. Rev. B **64** 092505

[24] Lorenz B, Li Z G, Honma T and Hor P H 2002 Phys. Rev. B **65** 144522

[25] Zhao Z X and Dong X L 1998 In Gap Symmetry and Fluctuations in High-T(C) Superconductors (J. Bok, G. Deutscher, D. Pavuna, and S. A. Wolf, eds.) **371** 171

[26] Dong X L, Dong Z F, Zhao B R, Zhao Z X, Duan X F, Peng L M, Huang W W, Xu B, Zhang Y Z, Guo S Q, Zhao L H and Li L 1998 Phys. Rev. Lett. **80** 2701

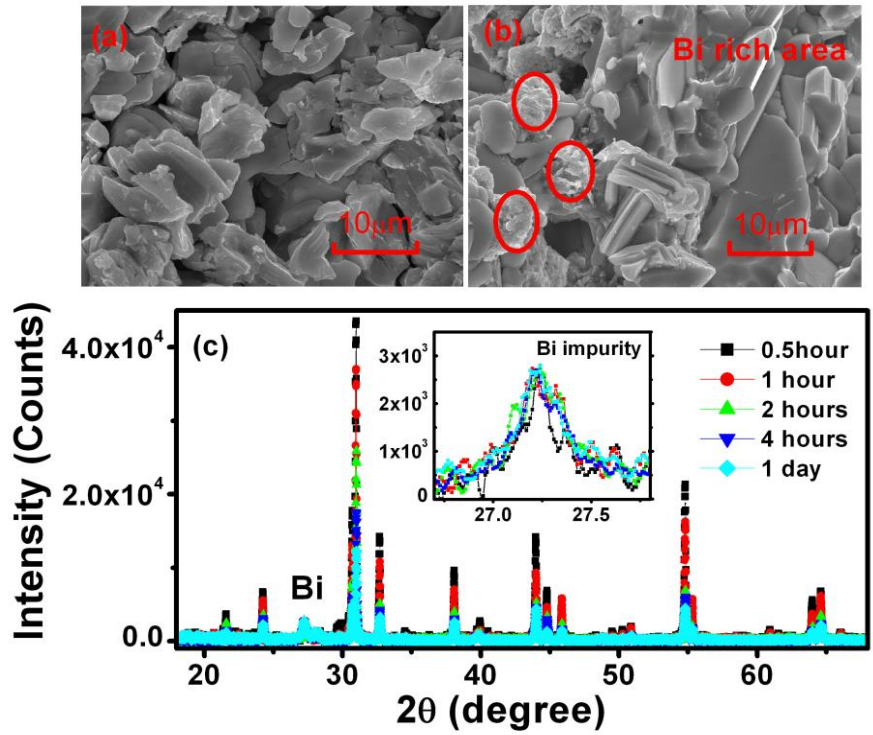

**Fig. 1** SEM images of BaTi$_2$(Sb$_{0.84}$Bi$_{0.16}$)$_2$O exposed to air for about one hour, with (a) representing to the most extent the typical texture features without Bi impurity and (b) the area where little Bi impurity can be detected in very limited local grain boundary, as indicated by the circles. (c) Powder X-ray diffraction patterns for samples with various exposure durations.

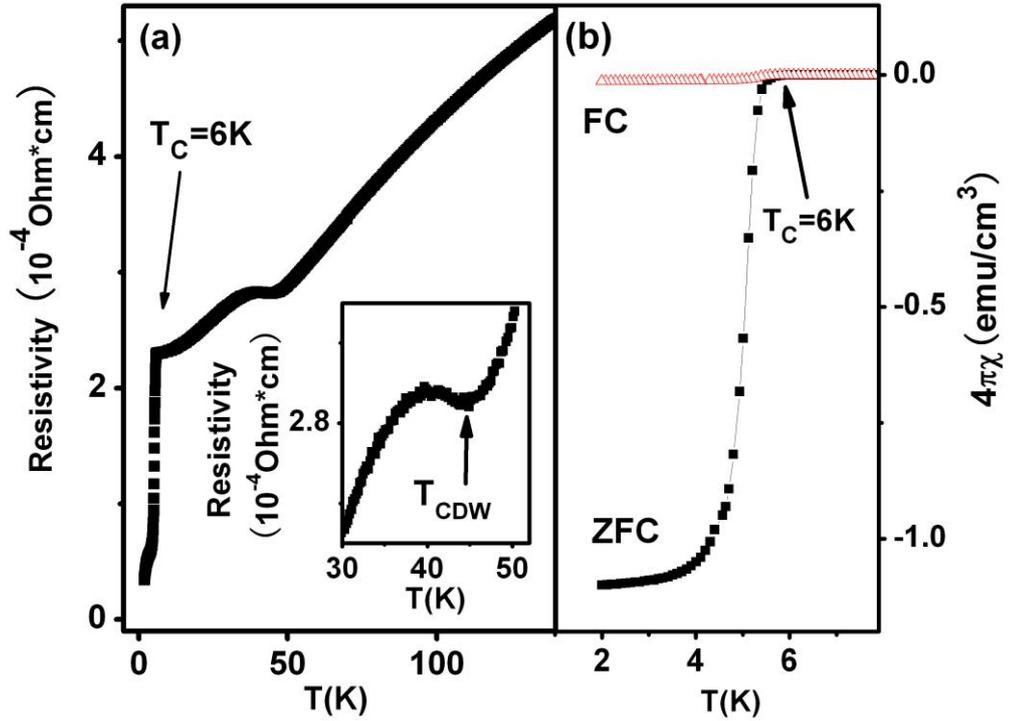

**Fig. 2** (a) Temperature dependence of resistivity for the as-prepared BaTi$_2$(Sb$_{0.84}$Bi$_{0.16}$)$_2$O. The expanded plot shown in the inset indicates the CDW/SDW anomaly below about 44K. (b) Temperature dependence of ZFC (solid squares) and FC (hollow triangles) magnetic susceptibilities of as-prepared BaTi$_2$(Sb$_{0.84}$Bi$_{0.16}$)$_2$O.

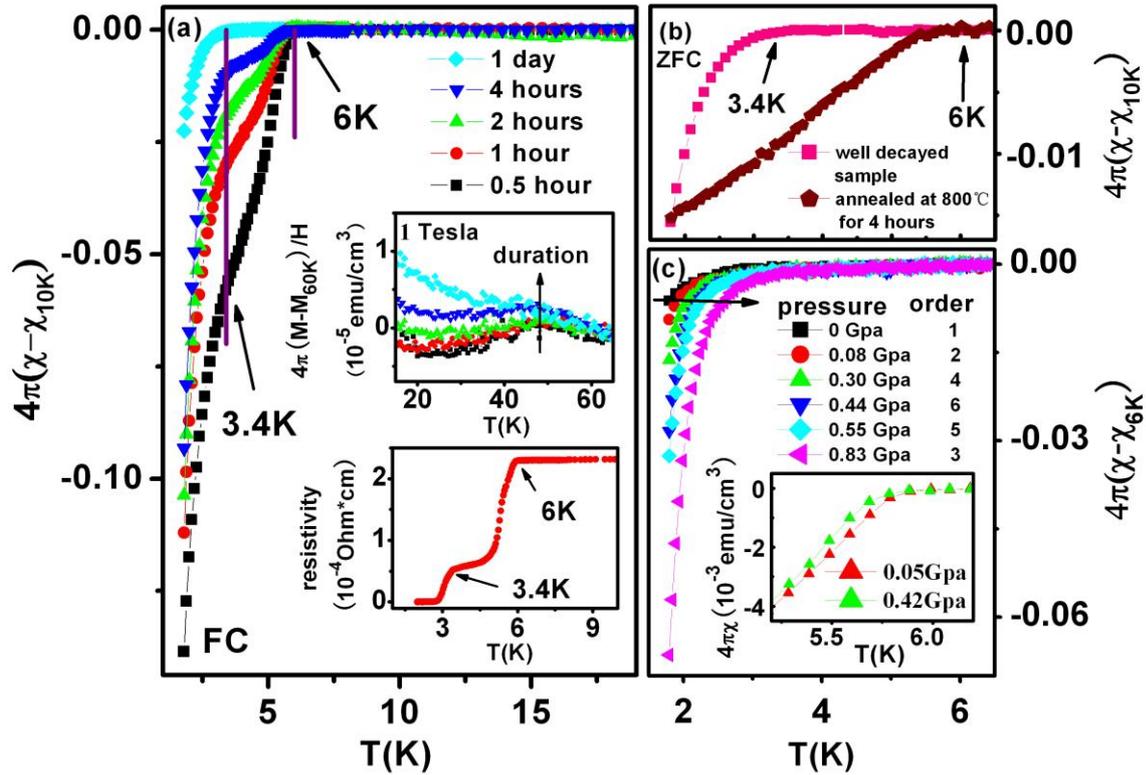

**Fig. 3** (a) Temperature dependences of FC magnetic susceptibility of BaTi$_2$(Sb$_{0.84}$Bi$_{0.16}$)$_2$O samples with various exposure durations. Shown in the upper inset is the CDW/SDW anomaly in magnetization for the same samples. The lower inset is the resistivity curve for the sample with one hour exposure, where two superconducting transitions can be clearly seen. (b) Temperature dependences of ZFC magnetic susceptibility for a well decayed sample and for the same sample annealed again at 800℃ for four hours. (c) FC magnetic susceptibility curves under pressures generated in a non-monotonic order for well decayed sample with low T$_C$ SC phase. Inset is the susceptibility data under pressures for the high T$_C$ SC phase for comparison. See text for details.

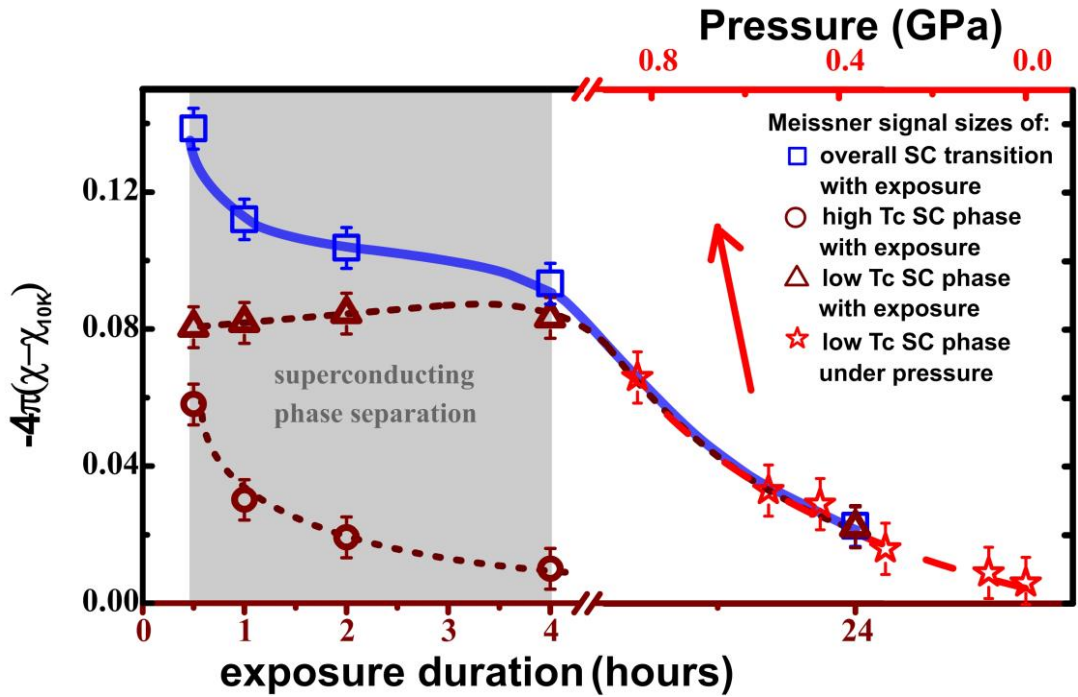

**Fig. 4** The Meissner signal sizes versus the exposure duration for the overall (hollow blue squares), the high $T_C$ (hollow brown circles) and the low $T_C$ (hollow brown triangles) superconducting transitions. Note a broken lower x-axis is used. Shown also in the right part is the pressure dependence of the Meissner signal size for the well decayed sample with low $T_C$ SC phase (hollow red stars). The Meissner signal sizes are taken here as the susceptibilities at 1.8 K in Fig. 3(a) for the overall SC transition and that in Fig. 3(c) for SC transition under pressure for well decayed sample. The light gray area marks the superconducting phase-separation regime, where the Meissner signal sizes allotted for the two coexisting high $T_C$ and low $T_C$ SC phases are defined as the changes of Meissner signal from 6 K to 3.4 K and from 3.4 K to 1.8 K, respectively. See text for details.